

\catcode`\@=11
\def\slash{\mathpalette\make@slash}
\def\make@slash#1#2{\setbox\z@\hbox{$#1#2$}%
  \hbox to 0pt{\hss$#1/$\hss\kern-\wd0}\box0}
\catcode`\@=12 

\magnification=\magstep1
\vsize=24truecm
\interlinepenalty=1000
\hsize=16.5truecm
\baselineskip=18pt
\hoffset=1truecm
\voffset=.5cm

 at 14.4 truept
\def\lam{\lambda}
\def\al{\alpha}
\def\pa{\partial}
\def\lap{\triangle}
\def\pr{\prime}
\def\ov{\over}
\def\ha{{1\over 2}}
\def\>{\rangle}
\def\<{\langle}
\def\es{\!=\!}

\def\mtx#1{\quad\hbox{{#1}}\quad}
\input lecproc.cmm

\contribution{Conformal- and Thermodynamic Properties of a Family of
Thirring-like Models}
\contributionrunning{Thirring-like models}
\author{I. Sachs@1, A. Wipf@1 and A. Dettki@2}
\address{@1Institute for Theoretical Physics, Eidgen\"ossische Technische
Hochschule, H\"onggerberg, CH-8093 Z\"urich, Switzerland
@2 Max-Planck Institut f{\"u}r Physik, Werner-Heisenberg
Institut f{\"u}r Physik, P.O. Box 40 12 12, Munich, Germany}
\topinsert\vbox{\hfill\hbox{ETH-TH/93-29}\smallskip\hrule}\endinsert
\abstract{We investigate Thirring-like models containing fermionic and
scalar fields propagating in 2-dimensional space
time. The corresponding conformal algebra is studied and we disprove
a conjecture relating the finite size effects to the central charge.
Some new results concerning the fermionic determinant on the torus
with chirally twisted boundary conditions and a chemical
potential are presented. In particular we show how the thermodynamics
of the Thirring model depends on the current-current interaction.}
The dependence
of expectation values on the temperature, particle density,
space region, imposed boundary conditions or external fields
is of importance in all the branches of physics [1]. In the present
work we address these
questions for the 2-dimensional
model defined by the action
$$
\eqalign{S=\int\sqrt{-g}\Big[&
i\bar\psi \gamma^\mu (\nabla_\mu- ig_1\pa_\mu\lam+ig_2
\eta_\mu^{\;\;\nu}\pa_\nu\phi )\psi\cr
&+g^{\mu\nu}(\pa_\mu\phi\pa_\nu\phi+\pa_\mu\lambda\pa_\nu\lam )-g_3
{\cal R}\lam\Big], \cr}\eqno(1)
$$
for the fermionic, scalar and pseudo
scalar fields $\psi,\lam$ and $\phi$ respectively. For $g_3\es 0$ and
$g_1^2\es-g_2^2\es g^2$
(1) belongs to the well known Thirring model
[2,3]. For $g_1\es g_2\es0$ the theory
decouples into free fields and non-minimally coupled
scalars describing the minimal models in CFT. $g_3$
measures the deviation from minimal coupling to gravitation.\par
On flat space-time (${\cal R}\es0$) the action (1) defines a
conformal field theory admitting a $U(1)$-Kac-Moody symmetry algebra. But
the conformal algebra is deformed relative to
that of the Thirring model. Part of
this deformation can already be seen on the classical level. Indeed
the energy momentum tensor
$$\eqalign{
T^{\mu\nu}
=\;&
{i\over 2}  \bigl [ \bar\psi\gamma^{(\mu}D^{\nu)}\psi -
(D^{(\mu}\bar\psi)\gamma^{\nu)}\psi\bigr ]\cr &
+2 \nabla^\mu\phi\nabla^\nu\phi -g^{\mu\nu}
\nabla^\alpha\phi\nabla_\alpha\phi
\quad + \quad (\phi \leftrightarrow \lam ) \cr &
-2 g_3 ( g^{\mu\nu}\nabla^2 - \nabla^{\mu}\nabla^{\nu})\lam \cr&
+\ha j^{\mu}\;(g_1\nabla^{\nu}\lam-g_2\eta^{\nu \al}\nabla_\al\phi
 ) \quad + \quad (\mu \leftrightarrow\nu)\cr
&+g_2 g^{\mu\nu} j^\al\eta_{\al\beta}
\nabla^\beta\phi-2g_2 j^\al\eta_\al\,^{(\mu}
\nabla^{\nu)}\phi \; ,
\cr}\eqno (2a)
$$
has trace
$$
T^\mu_{\,\,\,\mu}=g_3 ^2 {\cal{R}}.\eqno (2b)
$$
For $g_3\es 0$ the trace vanishes, and the theory becomes Weyl-invariant.
Hence it reduces to a conformal field theory in the
flat spacetime limit [4]. However, (1) can be made Weyl invariant even for
$g_3\!\neq\!0$ by adding a nonlocal Wess-Zumino-type
term to the action
$$
S\rightarrow S^\pr= S-{g_3^2\ov 4}S_p\mtx{where} S_p=\int \sqrt{-g}
{\cal{R}}{1\ov\lap}{\cal{R}}.
 \eqno (2c)
$$
The field equations are not affected by the WZ term
but the energy momentum tensor is modified in such a way that
its trace vanishes
and thus for $g_{\mu \nu}\! \to \! \eta _{\mu \nu}$ the Lagrangian
corresponds to a conformal field theory in Minkowski spacetime.
\par \noindent
The conformal weights of the fundamental fields are obtained
computing their poisson brackets [5] with the generator $T_f$ of the
conformal symmetry transformations. In light cone coordinates
$x^\pm=x^0\pm x^1,\;\;$
$T_f=\int dx^-f(x^-)T_{--}$
and
$$\eqalign{
\{T_f,\phi\}&=f\pa_-\phi\cr
\{T_f,\lam\}&=f\pa_-\lam-{g_3\ov 2}\pa_-f\cr
\{T_f,\psi_+\}&=f\pa_-\psi_++\ha(1-ig_1g_3)\psi_+\pa_-f\cr
\{T_f,\psi_+^{\dagger}\}&=f\pa_-\psi_+^{\dagger}+\ha(1+ig_1g_3)
\psi_+^{\dagger}\pa_-f,\cr}
\eqno(3)
$$
where $\psi_+\es\ha(1+\gamma_5)\psi$ denotes the right moving fermions.
$\phi$ and $\psi_+$ are primary fields with conformal weights
$h_{\phi}\es 0$ and $h_{{\psi_+}} \es\ha(1-ig_1g_3)$, respectively.
The non-primary character of $\lam$ is
linked with the $g_3$-dependent term in the transformation of the Dirac
field. Since $\psi$ is not a scalar under conformal transformation
the term $\bar\psi(...)\psi$ in (1) is only conformally invariant because
$\lam$ transforms inhomogenously like a spin connection.
It may be surprising that the symmetry transformations depend on the
coupling constant $g_3$ which is not present in the flat space time
Lagrangean. However, the same happens in $4$ dimensions
if one couples a scalar field conformally to
gravity. Although the Lagrangeans for the minimally and conformally
coupled particles are the same on Minkowski spacetime, their
energy momentum tensors are not. The same happens for the conformally
invariant nonabelian Toda theories wich admit several energy
momentum tensors and hence several conformal structures [6].
\par\noindent
The current transforms as
$$
\{T_f,j_-\}=f\pa_-j_-+j_-\pa_-f\eqno(4a)
$$
and hence is a primary field with weight $1$. For the energy momentum
tensor we find
$$
\{T_f,T_{--}\}=f\pa_-T_{--}+2T_{--}\pa_-f-g_3^2\pa_-^3f\eqno(4b)
$$
and thus a classical central charge $c\es 24\pi g_3^2$.
\par
The {\it quantum analogues} of (3,4) are encoded in the short distance
expansion of the fields with the stress tensor.
Stress tensor insertions are gotten by differentiation w.r.t. the metric as
$$\eqalign{
Z(g)\<O_1(x_1)...&O_n(x_n)T_{y_1y_1}...T_{y_ny_n}\>=\cr
&{(-2)^n\ov\sqrt{g(y_1)...g(y_n)}}{\delta^n\ov\delta g^{y_1y_1}...
\delta g^{y_ny_n}}\<O_1(x_1)...O_n(x_n)\>Z(g).\cr}\eqno(5)
$$
For a primary field $O$ with conformal weight $h$ one then finds the
transformation law
$$
{1\ov i}\oint dz f(z)
\<O(x) \; T_{zz}\> =
f(x)\pa_xO(x)+hO(x)\pa_xf(x).\eqno(6)
$$
Note that we
have switched to the Euclidian region for the quantum considerations.
The metric dependence of the effective action follows essentially from
the trace anomaly. One finds
$$
Z(g) = e^{({2\ov 24\pi}+g_3^2)S_L}Z(\hat g),\mtx{where}
g=e^{2\sigma}\hat g\eqno(7)
$$
and $S_L\es-\int\sqrt{g}\sigma\lap\sigma$ is the Liouville action.
Application of (5) and (6) then
yields
$$\eqalign{
&c=3+24 g_3^2 \pi\;\qquad\;h_j=1\cr
&h_{ {\psi_0}}=\;\ha+{1\ov  16 \pi} g_1^2
-{1\ov  16\pi}{2\pi g_2^2\ov  2\pi + g_2^2 }
-{ig_1 g_3 \ov  2} \cr
&h_{ {\psi_1^\dagger }}=\; (h_{ {\psi_0}})^{\dagger} \cr
&\bar h_{ {\psi_0}}=\;{1\ov  16 \pi} g_1^2
-{1\ov16\pi} {2\pi g_2^2\ov2\pi+g_2^2 }
-{ig_1 g_3 \ov  2}\;.\cr} \eqno(8)
$$
The classical results for the $\lambda$- and $\phi$ fields are
not modified. The central extension of the Kac-Moody
algebra and
the corresponding charges of the fermionic fields are the same as in
the original Thirring model [4]. In the Thirring model limit $g_3 \es 0$
and $g_1 \es g_2 \es g$,
the different contributions in (8) add up to give the
known anomalous dimension
appearing in the Thirring model [4]. The last classical term is a peculiar
feature of the solution.
For the conformal weights to be real we must choose an imaginary $g_3$.\par
Let us now quantize the system on a sphere. The presence of the length
scale breaks the conformal invariance and gives rise
to {\it finite size effects}. An effective method to compute finite
size effects has been
developped in [7]. It is based on the following
observation: Any conformal transformation $z\to w(z)$ is a composition
of a diffeomorphism (defined by the same $w$) and a compensating
Weyl transformation $g_{\mu\nu}\to e^{2\sigma}g_{\mu\nu}$
with
$$
e^{2\sigma}={dw(z)\ov dz}{d\bar w(\bar z)\ov d\bar z},\qquad
z=x^0+ix^1.\eqno(9)
$$
Therefore, chosing a diffeomorphism invariant regularization
one has for the effective action $\Gamma$
$$
0=\delta \Gamma_{Diff}=\delta\Gamma_{Conf}-\delta\Gamma_{Weyl}.\eqno(10)
$$
Integrating the conformal anomaly we end up with
$$
\delta\Gamma={g_3^2\ov 4}\int\sqrt{g}{\cal{R}}{1\ov\Delta} ({\cal{R}}-{8\pi\ov
V})
-{3\ov 24\pi }\int\sqrt{\hat g}\hat{\cal{R}}\sigma+{3\ov 24\pi}\int\sqrt{\hat
g}
\sigma\hat\Delta\sigma.\eqno(11)
$$
Now we can see why
the finite size conjecture generally fails to be true, although
it holds for theories without background charge on domains with boundaries
[7]. Take the simple case of a dilatation $w(z)\es a z$. Then, the
conformal angle is a constant $\sigma\es\log a$ and $({\cal{R}}-8\pi/V)\es 0$.
The first term in (11) vanishes and the finite
size effect does not depend on $g_3 ^2$. It is
given by
$$
\delta\Gamma=-{3\ov 24\pi}\log a\int\sqrt{\hat g}\hat{\cal{R}}=-\log a\eqno(12)
$$
and does not lead to the correct central charge $c$ in (8) which depends on
$g_3$. Thus we have disproved the conjecture. On other
Riemannian surfaces one would find similar results.
\par
To investigate the {\it thermodynamics}
of (1) we quantize the model on a flat torus [8]
with coordinates such that $x^\mu\in[0,L]$ and
$$
g_{\mu\nu}\equiv
\pmatrix{\vert\tau\vert^2 & \tau_1\cr \tau_1&1\cr},
$$
where $\tau\es\tau_1+i\tau_0$ is the Teichmueller parameter.
Furthermore we introduce a chemical potential for the conserved
$U(1)$-charge. Two
new features
appear which will have important consequences below:
\par\noindent
1.) In the euclidean theory the chemical potential is equivalent to a
constant imaginary gauge
potential [9]. Therefore one has to give a sensible definition for
fermionic determinants in complex gauge potentials.\par\noindent
2.) In order to recover the Thirring model on the torus one has to add
constant (harmonic) contributions $h_\mu$ to the auxillary field.
Hence  we complete the action (1) by adding $g_0{2\pi\over L}h_\mu
j^\mu+({2\pi\over L})^2
h_\mu h^\mu$, where $j^\mu\es \psi^\dagger\gamma^\mu\psi$, to the
Lagrangean in (1).\par\noindent
In order to compare our results with previous ones in the
literature [10,11] we allow for twisted boundary conditions for the
fermions
$$\eqalign{
\psi(x^0+L,x^1)&= -e^{2\pi i(\al_0+
\beta_0\gamma_5)}\psi(x^0,x^1)\cr
\psi(x^0,x^1+L)&=- e^{2\pi
i(\al_1+\beta_1\gamma_5)}\psi(x^0,x^1),\cr} \eqno(13)
$$
where $\gamma_5\es\sigma_3$. $\al_i$ and $\beta_i$
represent
vectorial and chiral twists, respectively. In fact, the chiral twists
are equivalent to chemical potentials. For the scalar field we
impose periodic boundary conditions. As a first step in
computing the {\it partition function} $Z$ of our model, we
determine the fermionic determinant. Due to the scaling
property
$$\eqalign{
&\slash D =\gamma^\nu D_\nu=e^{ig_1\lambda+g_2\gamma_5 \phi}
\,\slash {\hat D}\,e^{-ig_1\lambda+g_2\gamma_5 \phi},\mtx{where}\cr
&\slash {\hat D}=\gamma^\mu\big(\pa_\mu-{2\pi i\over
L}[g_0h_\mu+\mu_\mu]\big),\cr
&\mu_\mu=
-i{\tau_0L\over2\pi}\mu\;\delta_{\mu 0},\cr}\eqno(14)
$$
the dependence of ${\det} (i\slash D)$ on $\lam$ and $\phi$ can be
found integrating the
chiral anomaly [12] to be
$$
{\det} (i\slash D) = {\det} (i\slash {\hat D})
\exp\Big({1\ov 2\pi}\int\sqrt{g}
\phi\lap\phi\Big).\eqno(15)
$$
Using standard grassmann integration
rules [8,10] one further obtains
$$
\det i\slash {\hat D}=\prod_n\lam_n^+\lam_n^-,\eqno(16a)
$$
where
$$\eqalign{
\lam_n^+=& {2\pi\ov
\tau_0L}[\bar\tau(\ha+a_1+\beta_1+n_1)-(\ha+a_0+\beta_0+n_0)]\cr
\lam_n^-=& {2\pi\ov
\tau_0L}[\tau(\ha+a_1-\beta_1+n_1)-(\ha+a_0-\beta_0+n_0)],\cr
a_\mu=&\al_\mu-h_\mu-\mu_\mu.\cr}\eqno(16b)
$$
One may be tempted so identify
$$
\det (D_+D_-)\sim\prod\lambda_n^+\lambda_n^-\mtx{and}
\det D_+\det D_-\sim\prod\lambda_n^+\prod\lambda_m^-\eqno(16c)
$$
and thus conclude that the determinant is a product,
$f(\tau)\bar f(\tau)$, that is factorizes into holomorphic
and anti-holomorphic pieces. However, the infinite product (16a) must be
regularized and the two expressions in (16c) may differ.
To continue we recast the infinite product in the form
$$
\prod^\infty\lambda_n^+\lambda_n^-=\prod_{{\vec{n}}\in Z^2}
\Big({2\pi\ov L}\Big)^2 g^{\mu\nu}(\ha+c_\mu+n_\mu)(\ha+c_\nu+n_\nu)
\eqno(17a)$$
where
$$
c_\mu=a_\mu+i\eta_\mu\,^\nu\beta_\nu,\mtx{;}
(\eta_\mu\,^\nu)=-{1\ov\tau_0}\pmatrix{\tau_1&-|\tau|^2\cr 1&-\tau_1\cr}.
\eqno(17b)
$$
The point is that for real $c_\mu$, that is for vanishing chiral
twists $\beta_\mu$ and chemical potential the
zeta function defined by
$$
\zeta(s)=\sum_n\big(\lambda_n^+\lambda_n^-\big)^{-s}\eqno(17c)
$$
has a well defined analytic continuation to $s\!<\!1$ via
the Poisson resummation. However, for complex $c_\mu$ the Poisson resummation
is not applicable and $\zeta^\pr (0)$ cannot be calculated
by direct means. To circumvent these difficulties
we note that the infinite product (17c) defining the
$\zeta$-function for $s\!>\!1$ is a meromorphic function in $c$.
Thus we may first continue to $s\!<\!1$ for real $c_\mu$
and then continue the result to complex values. In this way we end up
with
$$
{\det} (i\slash {\hat D})={1\over |\eta(\tau)|^2}\Theta\Big[{-c_1\atop
c_0}\Big](0,\tau)
\bar\Theta\Big[{-\bar c_1\atop \bar c_0}\Big](0,\tau).\eqno(18)
$$
It can be
shown that this
determinant is {\it gauge invariant}, i.e. invariant under
$\al_\mu\rightarrow \al_\mu\!+\!1$, but not invariant under chiral
transformatins, $\beta_\mu\rightarrow \beta_\mu\!+\!1$, as expected.
Furthermore it transforms covariantly under modular
transformations $\tau\to\tau+1$ and $\tau\to -1/\tau$.
The result (18) also follows with operator methods and
differs from previous results in the literature [10]. In
particular there is no holomorphic factorization.\par
Having determined the fermionic determinant we are left with the
integration over the auxillary fields $\phi,\lambda$ and $h$. These
are Gaussian and one finds
$$
{Z\ov {\cal N}_0}={1\ov \vert\eta(\tau)\vert^2}
\sqrt{2\pi+g_2^2\ov 2\pi+g_0^2}\;
\Theta\Big[{u\atop w}\Big](\Lambda)\,,\eqno(19.a)
$$
where [13]
$$
\Theta\Big[{u\atop w}\Big](\Lambda)=
\sum\limits_{n\in Z^2}\,e^{i\pi(n+u)\Lambda (n+u)+2\pi i (n+u)w}\eqno(19.b)
$$
is the theta function with characteristics
$$
u=-\pmatrix{1\cr 1\cr}(\al_1+i\eta_1^{\;\nu}\beta_\nu)\mtx{and}
w=\pmatrix{1\cr -1\cr}(\al_0+i\eta_0^{\;\nu}\beta_\nu-\mu_0)\eqno(19.c)
$$
and covariance
$$
\Lambda=\pmatrix{\tau&0\cr 0&-\bar \tau\cr}
+i{\pi g_0^2\tau_0\ov 2\pi+g_0^2}\pmatrix{g_0^2&-4\pi-g_0^2\cr
-4\pi-g_0^2&g_0^2\cr}.\eqno(19.d)
$$
In (19.a) we have divided by the partition function ${\cal{N}}_0$ for
non-interacting auxillary fields, so the number of degrees of
freedom is the same as in the original Thirring model.
In the Thirring model limit $g_2\es g_0$ and the
square-root in (19a) disappears.\par
To investigate the thermodynamics of the model we must choose $\tau\es
i\beta/L$, where $\beta$ is the inverse temperature. Then
$$
\Omega=-{1\ov \beta}\log {Z\ov{\cal N}_0}\eqno(20)
$$
is the {\it grand canonical potential}. In the zero temperature limit
the saddle point approximation to the theta function in (19.a) becomes
exact. For {\it vanishing chiral twists and
chemical potential} this yields the the ground state energy
$$
E_0(L,\al_1,\beta_1\es 0)=-{\pi\ov 6L}+{2\pi\ov L}{2\pi\ov 2\pi+g_0^2}
\big(\al_1-[\ha+\al_1]\big)^2,\eqno(21)
$$
in agreement with [11]. Only for anti-periodic boundary conditions, that is for
$\al_1\es 0$,
does this Casimir energy coincide with the corresponding result for
free fermions. For $g_0^2\!\geq\!4\pi$ the Casimir force is always
attractive whereas for $g_0^2\!<\! 4\pi$ it can be attractive or
repulsive, depending on the value of $\al_1$.\par\noindent
For {\it small twists and chemical potential} the grand potential
becomes
$$
\Omega(\beta\to\infty)=-{\pi\ov 6L}+{2\pi\ov L}{2\pi\ov 2\pi+g_0^2}
\al_1^2,\eqno(22)
$$
and hence, is independent of chemical potential and chiral twist. Here we
disagree with [11]. The discrepancy is due to the breakdown of
holomorphic factorization, a property which
has been presupposed in [11]. In order to show that the result (22) is
physically reasonable we note that for massless fermions the Fermi energy is
just $\mu$ and
at $T\es 0$ all electron states with energies less then $\mu$
and all positron states with energies less then $-\mu$ are filled.
The other states are empty. Since $d\Omega/d\mu$
is the expectation value of the electric charge in the
presence of $\mu$ we conclude that it must jump if $\mu$ crosses
an eigenvalue of the first quantized Dirac Hamiltonian $h$.
For vanishing twists the eigenvalues of $h$ are just $E_n\es(n-\ha)\pi/L $.
Indeed, from (22) one finds that the electric charge
$$
\<Q\>={d\Omega \ov d\mu}=2
\big[\ha+{\mu L\ov 2\pi}\big]=2n\quad\mtx{for} E_n\leq\mu<E_{n+1}\eqno(23.a)
$$
jumps at these values for $\mu$. Further observe,
that in the {\it thermodynamic limit} $L\to\infty$ the density
$$
{\Omega\ov L}\rightarrow - {2\pi\ov 2\pi+g_0^2}{\mu^2\ov 2\pi}, \eqno(23.b)
$$
reduces for $g_0\es 0$ to the standard result for free electrons.\par
Let us now discuss the {\it equation of state}. Using the
transformation properties of the theta functions under modular
transformations [13] the pressure is given for $L\to\infty$ and
{\it small twists} by
$$
\beta p={\pi\ov 6\beta}-{2\pi\ov\beta}{2\pi\ov 2\pi+g_0^2}
\big(\al_0+i{\beta\mu\ov 2\pi}\big)^2.\eqno(24.a)
$$
In particular it becomes independent on the chiral twist
$\beta_0$ in agreement with
the earlier result that for small twists $\Omega$ is independent of
$\beta _1$. For the thermal boundary conditions $\al_0\es
0$, we are lead
to the following equation of state
$$
p(\beta,\mu,\al_0\es 0) ={\pi\ov 6\beta^2}+
{\mu^2\ov 2\pi}{2\pi\ov 2\pi+g_0^2},\eqno(24.b)
$$
which for small $\beta_0$ relates the pressure to the chemical
potential and temperature.
This result is consistent with the renormalization of the
electric charge which is conjugate to the chemical potential.
It shows in particular that the thermodynamic
behaviour of the Thirring model is
not just the one of free fermions as has been claimed in
[14]. Indeed, the zero point pressure is multiplied by a
factor $2\pi/(2\pi+g_0^2)$.
This modification arises from the coupling of the current to the
harmonic fields. It can not be seen if only the local part of the
auxillary field is considered, which is the case if one quantizes the
model on the infinite Euclidean space.\par\bigskip

This work has been supported by the Swiss National Science
Foundation. We would like to thank K. Gawedzki, C. Kiefer and E. Seiler
for discussions.

\begrefchapter{References}
\refno{1.} W. Dittrich and M. Reuter, Effective Lagrangians in QED,
Lecture notes in Physics, Springer, Heidelberg, 1984.
\refno{2.} B. Klaiber, Lecture notes in Phys. XA, Gordon and
Breach, New York, 1968.
\refno{3.} A.J. da Silva, M. Gomes and R. K{\"o}berle, Phys. Rev.
{\bf D34} (1986) 504; M. Gomes and A.J. da Silva, Phys. Rev. {\bf D34}
(1986) 3916.
\refno{4.} P. Furlan, G.M. Sotkov and I.T. Todorov, Riv. Nuovo Cim.
{\bf 12} (1989) 1.
\refno{5.} R. Casalbuoni, Il Nuovo Cim. {\bf A33} (1976) 115.
\refno{6.} L. O'Raifeartaigh and A. Wipf, Phys. Lett. {251B} (1990) 361;
L. Feher, L. O'Raifeartaigh, P. Ruelle, I. Tsutsui and A. Wipf,
Phys. Rep. {\bf 222} (1992) 1.
\refno{7.} A. Dettki and A. Wipf, Nucl. Phys. {\bf B377} (1992) 252.
\refno{8.} I. Sachs and A. Wipf, Helv. Phys. Acta {\bf 65} (1992) 653.
\refno{9.} A. Actor, Fortschritte der Phys. {\bf 35} (1987) 793;
\refno{10.} D.Z. Freedman and K. Pilch, Phys. Lett. {\bf 213B} (1988) 331;
D.Z. Freedman and K. Pilch, Ann. Phys. {\bf 192} (1989) 331;
S. Wu, Comm. Math. Phys. {\bf 124} (1989) 133.
\refno{11.} C. Destri and J.J. deVega, Phys. Lett. {\bf 223B} (1989) 365.
\refno{12.} S. Blau, M. Visser and A. Wipf, Int. J. Mod. Phys. {\bf A4}
(1989) 1467.
\refno{13.} D. Mumford, Tata Lectures on Theta, Birkh\"auser, Boston 1983.
\refno{14.} H. Yokota, Prog. Theor. Phys. {\bf 77} (1987) 1450.

\end

\magnification=\magstep1
\font \authfont               = cmr10 scaled\magstep4
\font \fivesans               = cmss10 at 5pt
\font \headfont               = cmbx12 scaled\magstep4
\font \markfont               = cmr10 scaled\magstep1
\font \ninebf                 = cmbx9
\font \ninei                  = cmmi9
\font \nineit                 = cmti9
\font \ninerm                 = cmr9
\font \ninesans               = cmss10 at 9pt
\font \ninesl                 = cmsl9
\font \ninesy                 = cmsy9
\font \ninett                 = cmtt9
\font \sevensans              = cmss10 at 7pt
\font \sixbf                  = cmbx6
\font \sixi                   = cmmi6
\font \sixrm                  = cmr6
\font \sixsans                = cmss10 at 6pt
\font \sixsy                  = cmsy6
\font \smallescriptfont       = cmr5 at 7pt
\font \smallescriptscriptfont = cmr5
\font \smalletextfont         = cmr5 at 10pt
\font \subhfont               = cmr10 scaled\magstep4
\font \tafonts                = cmbx7  scaled\magstep2
\font \tafontss               = cmbx5  scaled\magstep2
\font \tafontt                = cmbx10 scaled\magstep2
\font \tams                   = cmmib10
\font \tamss                  = cmmib10
\font \tamt                   = cmmib10 scaled\magstep2
\font \tass                   = cmsy7  scaled\magstep2
\font \tasss                  = cmsy5  scaled\magstep2
\font \tast                   = cmsy10 scaled\magstep2
\font \tasys                  = cmex10 scaled\magstep1
\font \tasyt                  = cmex10 scaled\magstep2
\font \tbfonts                = cmbx7  scaled\magstep1
\font \tbfontss               = cmbx5  scaled\magstep1
\font \tbfontt                = cmbx10 scaled\magstep1
\font \tbms                   = cmmib10 scaled 833
\font \tbmss                  = cmmib10 scaled 600
\font \tbmt                   = cmmib10 scaled\magstep1
\font \tbss                   = cmsy7  scaled\magstep1
\font \tbsss                  = cmsy5  scaled\magstep1
\font \tbst                   = cmsy10 scaled\magstep1
\font \tenbfne                = cmb10
\font \tensans                = cmss10
\font \tpfonts                = cmbx7  scaled\magstep3
\font \tpfontss               = cmbx5  scaled\magstep3
\font \tpfontt                = cmbx10 scaled\magstep3
\font \tpmt                   = cmmib10 scaled\magstep3
\font \tpss                   = cmsy7  scaled\magstep3
\font \tpsss                  = cmsy5  scaled\magstep3
\font \tpst                   = cmsy10 scaled\magstep3
\font \tpsyt                  = cmex10 scaled\magstep3
\vsize=22.5true cm
\hsize=13.8true cm
\hfuzz=2pt
\tolerance=500
\abovedisplayskip=3 mm plus6pt minus 4pt
\belowdisplayskip=3 mm plus6pt minus 4pt
\abovedisplayshortskip=0mm plus6pt minus 2pt
\belowdisplayshortskip=2 mm plus4pt minus 4pt
\predisplaypenalty=0
\clubpenalty=10000
\widowpenalty=10000
\frenchspacing
\newdimen\oldparindent\oldparindent=1.5em
\parindent=1.5em
\skewchar\ninei='177 \skewchar\sixi='177
\skewchar\ninesy='60 \skewchar\sixsy='60
\hyphenchar\ninett=-1
\def\newline{\hfil\break}%
\catcode`@=11
\def\folio{\ifnum\pageno<\z@
\uppercase\expandafter{\romannumeral-\pageno}%
\else\number\pageno \fi}
\catcode`@=12 
  \mathchardef\Gamma="0100
  \mathchardef\Delta="0101
  \mathchardef\Theta="0102
  \mathchardef\Lambda="0103
  \mathchardef\Xi="0104
  \mathchardef\Pi="0105
  \mathchardef\Sigma="0106
  \mathchardef\Upsilon="0107
  \mathchardef\Phi="0108
  \mathchardef\Psi="0109
  \mathchardef\Omega="010A
  \mathchardef\bfGamma="0\the\bffam 00
  \mathchardef\bfDelta="0\the\bffam 01
  \mathchardef\bfTheta="0\the\bffam 02
  \mathchardef\bfLambda="0\the\bffam 03
  \mathchardef\bfXi="0\the\bffam 04
  \mathchardef\bfPi="0\the\bffam 05
  \mathchardef\bfSigma="0\the\bffam 06
  \mathchardef\bfUpsilon="0\the\bffam 07
  \mathchardef\bfPhi="0\the\bffam 08
  \mathchardef\bfPsi="0\the\bffam 09
  \mathchardef\bfOmega="0\the\bffam 0A

\def\sq{\hbox{\rlap{$\sqcap$}$\sqcup$}}

\def\utw{\smash{\rlap{\lower5pt\hbox{$\sim$}}}}
\def\udtw{\smash{\rlap{\lower6pt\hbox{$\approx$}}}}

\def\diameter{{\ifmmode\mathchoice
{\ooalign{\hfil\hbox{$\displaystyle/$}\hfil\crcr
{\hbox{$\displaystyle\mathchar"20D$}}}}
{\ooalign{\hfil\hbox{$\textstyle/$}\hfil\crcr
{\hbox{$\textstyle\mathchar"20D$}}}}
{\ooalign{\hfil\hbox{$\scriptstyle/$}\hfil\crcr
{\hbox{$\scriptstyle\mathchar"20D$}}}}
{\ooalign{\hfil\hbox{$\scriptscriptstyle/$}\hfil\crcr
{\hbox{$\scriptscriptstyle\mathchar"20D$}}}}
\else{\ooalign{\hfil/\hfil\crcr\mathhexbox20D}}%
\fi}}


\def\bbbc{{\mathchoice {\setbox0=\hbox{$\displaystyle\rm C$}\hbox{\hbox
to0pt{\kern0.4\wd0\vrule height0.9\ht0\hss}\box0}}
{\setbox0=\hbox{$\textstyle\rm C$}\hbox{\hbox
to0pt{\kern0.4\wd0\vrule height0.9\ht0\hss}\box0}}
{\setbox0=\hbox{$\scriptstyle\rm C$}\hbox{\hbox
to0pt{\kern0.4\wd0\vrule height0.9\ht0\hss}\box0}}
{\setbox0=\hbox{$\scriptscriptstyle\rm C$}\hbox{\hbox
to0pt{\kern0.4\wd0\vrule height0.9\ht0\hss}\box0}}}}
\def\bbbe{{\mathchoice {\setbox0=\hbox{\smalletextfont e}\hbox{\raise
0.1\ht0\hbox to0pt{\kern0.4\wd0\vrule width0.3pt height0.7\ht0\hss}\box0}}
{\setbox0=\hbox{\smalletextfont e}\hbox{\raise
0.1\ht0\hbox to0pt{\kern0.4\wd0\vrule width0.3pt height0.7\ht0\hss}\box0}}
{\setbox0=\hbox{\smallescriptfont e}\hbox{\raise
0.1\ht0\hbox to0pt{\kern0.5\wd0\vrule width0.2pt height0.7\ht0\hss}\box0}}
{\setbox0=\hbox{\smallescriptscriptfont e}\hbox{\raise
0.1\ht0\hbox to0pt{\kern0.4\wd0\vrule width0.2pt height0.7\ht0\hss}\box0}}}}
\def\bbbq{{\mathchoice {\setbox0=\hbox{$\displaystyle\rm Q$}\hbox{\raise
0.15\ht0\hbox to0pt{\kern0.4\wd0\vrule height0.8\ht0\hss}\box0}}
{\setbox0=\hbox{$\textstyle\rm Q$}\hbox{\raise
0.15\ht0\hbox to0pt{\kern0.4\wd0\vrule height0.8\ht0\hss}\box0}}
{\setbox0=\hbox{$\scriptstyle\rm Q$}\hbox{\raise
0.15\ht0\hbox to0pt{\kern0.4\wd0\vrule height0.7\ht0\hss}\box0}}
{\setbox0=\hbox{$\scriptscriptstyle\rm Q$}\hbox{\raise
0.15\ht0\hbox to0pt{\kern0.4\wd0\vrule height0.7\ht0\hss}\box0}}}}
\def\bbbt{{\mathchoice {\setbox0=\hbox{$\displaystyle\rm
T$}\hbox{\hbox to0pt{\kern0.3\wd0\vrule height0.9\ht0\hss}\box0}}
{\setbox0=\hbox{$\textstyle\rm T$}\hbox{\hbox
to0pt{\kern0.3\wd0\vrule height0.9\ht0\hss}\box0}}
{\setbox0=\hbox{$\scriptstyle\rm T$}\hbox{\hbox
to0pt{\kern0.3\wd0\vrule height0.9\ht0\hss}\box0}}
{\setbox0=\hbox{$\scriptscriptstyle\rm T$}\hbox{\hbox
to0pt{\kern0.3\wd0\vrule height0.9\ht0\hss}\box0}}}}
\def\bbbs{{\mathchoice
{\setbox0=\hbox{$\displaystyle     \rm S$}\hbox{\raise0.5\ht0\hbox
to0pt{\kern0.35\wd0\vrule height0.45\ht0\hss}\hbox
to0pt{\kern0.55\wd0\vrule height0.5\ht0\hss}\box0}}
{\setbox0=\hbox{$\textstyle        \rm S$}\hbox{\raise0.5\ht0\hbox
to0pt{\kern0.35\wd0\vrule height0.45\ht0\hss}\hbox
to0pt{\kern0.55\wd0\vrule height0.5\ht0\hss}\box0}}
{\setbox0=\hbox{$\scriptstyle      \rm S$}\hbox{\raise0.5\ht0\hbox
to0pt{\kern0.35\wd0\vrule height0.45\ht0\hss}\raise0.05\ht0\hbox
to0pt{\kern0.5\wd0\vrule height0.45\ht0\hss}\box0}}
{\setbox0=\hbox{$\scriptscriptstyle\rm S$}\hbox{\raise0.5\ht0\hbox
to0pt{\kern0.4\wd0\vrule height0.45\ht0\hss}\raise0.05\ht0\hbox
to0pt{\kern0.55\wd0\vrule height0.45\ht0\hss}\box0}}}}
\def\bbbz{{\mathchoice {\hbox{$\sans\textstyle Z\kern-0.4em Z$}}
{\hbox{$\sans\textstyle Z\kern-0.4em Z$}}
{\hbox{$\sans\scriptstyle Z\kern-0.3em Z$}}
{\hbox{$\sans\scriptscriptstyle Z\kern-0.2em Z$}}}}
\def\qed{\ifmmode\sq\else{\unskip\nobreak\hfil
\penalty50\hskip1em\null\nobreak\hfil\sq
\parfillskip=0pt\finalhyphendemerits=0\endgraf}\fi}
\newfam\sansfam
\textfont\sansfam=\tensans\scriptfont\sansfam=\sevensans
\scriptscriptfont\sansfam=\fivesans
\def\sans{\fam\sansfam\tensans}
\def\stackfigbox{\if
Y\FIG\global\setbox\figbox=\vbox{\unvbox\figbox\box1}%
\else\global\setbox\figbox=\vbox{\box1}\global\let\FIG=Y\fi}
\def\placefigure{\dimen0=\ht1\advance\dimen0by\dp1
\advance\dimen0by5\baselineskip
\advance\dimen0by0.4true cm
\ifdim\dimen0>\vsize\pageinsert\box1\vfill\endinsert
\else
\if Y\FIG\stackfigbox\else
\dimen0=\pagetotal\ifdim\dimen0<\pagegoal
\advance\dimen0by\ht1\advance\dimen0by\dp1\advance\dimen0by1.4true cm
\ifdim\dimen0>\pagegoal\stackfigbox
\else\box1\vskip4true mm\fi
\else\box1\vskip4true mm\fi\fi\fi}
%
\def\begfig#1cm#2\endfig{\par
\setbox1=\vbox{\dimen0=#1true cm\advance\dimen0
by1true cm\kern\dimen0#2}\placefigure}
\def\begdoublefig#1cm #2 #3 \enddoublefig{\begfig#1cm%
\vskip-.8333\baselineskip\line{\vtop{\hsize=0.46\hsize#2}\hfill
\vtop{\hsize=0.46\hsize#3}}\endfig}
\def\begfigsidebottom#1cm#2cm#3\endfigsidebottom{\dimen0=#2true cm
\ifdim\dimen0<0.4\hsize\message{Room for legend to narrow;
begfigsidebottom changed to begfig}\begfig#1cm#3\endfig\else
\par\def\figure##1##2{\vbox{\noindent\petit{\bf
Fig.\ts##1\unskip.\ }\ignorespaces ##2\par}}%
\dimen0=\hsize\advance\dimen0 by-.8true cm\advance\dimen0 by-#2true cm
\setbox1=\vbox{\hbox{\hbox to\dimen0{\vrule height#1true cm\hrulefill}%
\kern.8true cm\vbox{\hsize=#2true cm#3}}}\placefigure\fi}
\def\begfigsidetop#1cm#2cm#3\endfigsidetop{\dimen0=#2true cm
\ifdim\dimen0<0.4\hsize\message{Room for legend to narrow; begfigsidetop
changed to begfig}\begfig#1cm#3\endfig\else
\par\def\figure##1##2{\vbox{\noindent\petit{\bf
Fig.\ts##1\unskip.\ }\ignorespaces ##2\par}}%
\dimen0=\hsize\advance\dimen0 by-.8true cm\advance\dimen0 by-#2true cm
\setbox1=\vbox{\hbox{\hbox to\dimen0{\vrule height#1true cm\hrulefill}%
\kern.8true cm\vbox to#1true cm{\hsize=#2 true cm#3\vfill
}}}\placefigure\fi}
\def\figure#1#2{\vskip1true cm\setbox0=\vbox{\noindent\petit{\bf
Fig.\ts#1\unskip.\ }\ignorespaces #2\smallskip
\count255=0\global\advance\count255by\prevgraf}%
\ifnum\count255>1\box0\else
\centerline{\petit{\bf Fig.\ts#1\unskip.\
}\ignorespaces#2}\smallskip\fi}

\def\begtab#1cm#2\endtab{\par
   \ifvoid\topins\midinsert\medskip\vbox{#2\kern#1true cm}\endinsert
   \else\topinsert\vbox{#2\kern#1true cm}\endinsert\fi}
\def\begpet{\vskip6pt\bgroup\petit}
\def\endpet{\vskip6pt\egroup}
\newcount\frpages
\newcount\frpagegoal
\def\freepage#1{\global\frpagegoal=#1\relax\global\frpages=0\relax
\loop\global\advance\frpages by 1\relax
\message{Doing freepage \the\frpages\space of
\the\frpagegoal}\null\vfill\eject
\ifnum\frpagegoal>\frpages\repeat}
\newdimen\refindent
\def\begrefchapter#1{\titlea{}{\ignorespaces#1}%
\bgroup\petit
\setbox0=\hbox{1000.\enspace}\refindent=\wd0}
\def\ref{\goodbreak
\hangindent\oldparindent\hangafter=1
\noindent\ignorespaces}
\def\refno#1{\goodbreak
\hangindent\refindent\hangafter=1
\noindent\hbox to\refindent{#1\hss}\ignorespaces}
\def\vec#1{{\textfont1=\tams\scriptfont1=\tamss
\textfont0=\tenbf\scriptfont0=\sevenbf
\mathchoice{\hbox{$\displaystyle#1$}}{\hbox{$\textstyle#1$}}
{\hbox{$\scriptstyle#1$}}{\hbox{$\scriptscriptstyle#1$}}}}
\def\petit{\def\rm{\fam0\ninerm}%
\textfont0=\ninerm \scriptfont0=\sixrm \scriptscriptfont0=\fiverm
 \textfont1=\ninei \scriptfont1=\sixi \scriptscriptfont1=\fivei
 \textfont2=\ninesy \scriptfont2=\sixsy \scriptscriptfont2=\fivesy
 \def\it{\fam\itfam\nineit}%
 \textfont\itfam=\nineit
 \def\sl{\fam\slfam\ninesl}%
 \textfont\slfam=\ninesl
 \def\bf{\fam\bffam\ninebf}%
 \textfont\bffam=\ninebf \scriptfont\bffam=\sixbf
 \scriptscriptfont\bffam=\fivebf
 \def\sans{\fam\sansfam\ninesans}%
 \textfont\sansfam=\ninesans \scriptfont\sansfam=\sixsans
 \scriptscriptfont\sansfam=\fivesans
 \def\tt{\fam\ttfam\ninett}%
 \textfont\ttfam=\ninett
 \normalbaselineskip=11pt
 \setbox\strutbox=\hbox{\vrule height7pt depth2pt width0pt}%
 \normalbaselines\rm
\def\vec##1{{\textfont1=\tbms\scriptfont1=\tbmss
\textfont0=\ninebf\scriptfont0=\sixbf
\mathchoice{\hbox{$\displaystyle##1$}}{\hbox{$\textstyle##1$}}
{\hbox{$\scriptstyle##1$}}{\hbox{$\scriptscriptstyle##1$}}}}}
\nopagenumbers
%
\let\header=Y
\let\FIG=N
\newbox\figbox
\output={\if N\header\headline={\hfil}\fi\plainoutput\global\let\header=Y
\if Y\FIG\topinsert\unvbox\figbox\endinsert\global\let\FIG=N\fi}
\let\lasttitle=N
\def\bookauthor#1{\vfill\eject
     \bgroup
     \baselineskip=22pt
     \lineskip=0pt
     \pretolerance=10000
     \authfont
     \rightskip 0pt plus 6em
     \centerpar{#1}\vskip2true cm\egroup}
\def\bookhead#1#2{\bgroup
     \baselineskip=36pt
     \lineskip=0pt
     \pretolerance=10000
     \headfont
     \rightskip 0pt plus 6em
     \centerpar{#1}\vskip1true cm
     \baselineskip=22pt
     \subhfont\centerpar{#2}\vfill
     \parindent=0pt
     \baselineskip=16pt
     \leftskip=2.2true cm
     \markfont Springer-Verlag\newline
     Berlin Heidelberg New York\newline
     London Paris Tokyo Singapore\bigskip\bigskip
     [{\it This is page III of your manuscript and will be reset by
     Springer.}]
     \egroup\let\header=N\eject}
\def\centerpar#1{{\parfillskip=0pt
\rightskip=0pt plus 1fil
\leftskip=0pt plus 1fil
\advance\leftskip by\oldparindent
\advance\rightskip by\oldparindent
\def\newline{\break}%
\noindent\ignorespaces#1\par}}
\def\part#1#2{\vfill\supereject\let\header=N
\centerline{\subhfont#1}%
\vskip75pt
     \bgroup
\textfont0=\tpfontt \scriptfont0=\tpfonts \scriptscriptfont0=\tpfontss
\textfont1=\tpmt \scriptfont1=\tbmt \scriptscriptfont1=\tams
\textfont2=\tpst \scriptfont2=\tpss \scriptscriptfont2=\tpsss
\textfont3=\tpsyt \scriptfont3=\tasys \scriptscriptfont3=\tenex
     \baselineskip=20pt
     \lineskip=0pt
     \pretolerance=10000
     \tpfontt
     \centerpar{#2}
     \vfill\eject\egroup\ignorespaces}
\newtoks\AUTHOR
\newtoks\HEAD
\catcode`\@=\active
\def\author#1{\bgroup
\baselineskip=22pt
\lineskip=0pt
\pretolerance=10000
\markfont
\centerpar{#1}\bigskip\egroup
{\def@##1{}%
\setbox0=\hbox{\petit\kern2.5true cc\ignorespaces#1\unskip}%
\ifdim\wd0>\hsize
\message{The names of the authors exceed the headline, please use a }%
\message{short form with AUTHORRUNNING}\gdef\leftheadline{%
\hbox to2.5true cc{\folio\hfil}AUTHORS suppressed due to excessive
length\hfil}%
\global\AUTHOR={AUTHORS were to long}\else
\xdef\leftheadline{\hbox to2.5true
cc{\noexpand\folio\hfil}\ignorespaces#1\hfill}%
\global\AUTHOR={\def@##1{}\ignorespaces#1\unskip}\fi
}\let\INS=E}
\def\address#1{\bgroup
\centerpar{#1}\bigskip\egroup
\catcode`\@=12
\vskip2cm\noindent\ignorespaces}
\let\INS=N%
\def@#1{\if N\INS\unskip\ $^{#1}$\else\if
E\INS\noindent$^{#1}$\let\INS=Y\ignorespaces
\else\par
\noindent$^{#1}$\ignorespaces\fi\fi}%
\catcode`\@=12
\headline={\petit\def\newline{ }\def\fonote#1{}\ifodd\pageno
\rightheadline\else\leftheadline\fi}
\def\rightheadline{\hfil Missing CONTRIBUTION
title\hbox to2.5true cc{\hfil\folio}}
\def\leftheadline{\hbox to2.5true cc{\folio\hfil}Missing name(s) of the
author(s)\hfil}
\nopagenumbers
\let\header=Y
\def\contributionrunning#1{\message{Running head on right hand sides
(CONTRIBUTION)
has been changed}\gdef\rightheadline{\hfill\ignorespaces#1\unskip
\hbox to2.5true cc{\hfil\folio}}\global\HEAD={\ignorespaces#1\unskip}}

\let\lasttitle=N
 \def\contribution#1{\vfill\supereject
 \ifodd\pageno\else\null\vfill\supereject\fi
 \let\header=N\bgroup
 \textfont0=\tafontt \scriptfont0=\tafonts \scriptscriptfont0=\tafontss
 \textfont1=\tamt \scriptfont1=\tams \scriptscriptfont1=\tams
 \textfont2=\tast \scriptfont2=\tass \scriptscriptfont2=\tasss
 \par\baselineskip=16pt
     \lineskip=16pt
     \tafontt
     \raggedright
     \pretolerance=10000
     \noindent
     \centerpar{\ignorespaces#1}%
     \vskip12pt\egroup
     \nobreak
     \parindent=0pt
     \everypar={\global\parindent=1.5em
     \global\let\lasttitle=N\global\everypar={}}%
     \global\let\lasttitle=A%
     \setbox0=\hbox{\petit\def\newline{ }\def\fonote##1{}\kern2.5true
     cc\ignorespaces#1}\ifdim\wd0>\hsize
     \message{Your CONTRIBUTIONtitle exceeds the headline,
please use a short form
with CONTRIBUTIONRUNNING}\gdef\rightheadline{\hfil CONTRIBUTION title
suppressed due to excessive length\hbox to2.5true cc{\hfil\folio}}%
\global\HEAD={HEAD was to long}\else
\gdef\rightheadline{\hfill\ignorespaces#1\unskip\hbox to2.5true
cc{\hfil\folio}}\global\HEAD={\ignorespaces#1\unskip}\fi
\catcode`\@=\active
     \ignorespaces}
 \def\contributionnext#1{\vfill\supereject
 \let\header=N\bgroup
 \textfont0=\tafontt \scriptfont0=\tafonts \scriptscriptfont0=\tafontss
 \textfont1=\tamt \scriptfont1=\tams \scriptscriptfont1=\tams
 \textfont2=\tast \scriptfont2=\tass \scriptscriptfont2=\tasss
 \par\baselineskip=16pt
     \lineskip=16pt
     \tafontt
     \raggedright
     \pretolerance=10000
     \noindent
     \centerpar{\ignorespaces#1}%
     \vskip12pt\egroup
     \nobreak
     \parindent=0pt
     \everypar={\global\parindent=1.5em
     \global\let\lasttitle=N\global\everypar={}}%
     \global\let\lasttitle=A%
     \setbox0=\hbox{\petit\def\newline{ }\def\fonote##1{}\kern2.5true
     cc\ignorespaces#1}\ifdim\wd0>\hsize
     \message{Your CONTRIBUTIONtitle exceeds the headline,
please use a short form
with CONTRIBUTIONRUNNING}\gdef\rightheadline{\hfil CONTRIBUTION title
suppressed due to excessive length\hbox to2.5true cc{\hfil\folio}}%
\global\HEAD={HEAD was to long}\else
\gdef\rightheadline{\hfill\ignorespaces#1\unskip\hbox to2.5true
cc{\hfil\folio}}\global\HEAD={\ignorespaces#1\unskip}\fi
\catcode`\@=\active
     \ignorespaces}
\def\motto#1#2{\bgroup\petit\leftskip=6.5true cm\noindent\ignorespaces#1
\if!#2!\else\medskip\noindent\it\ignorespaces#2\fi\bigskip\egroup
\let\lasttitle=M
\parindent=0pt
\everypar={\global\parindent=\oldparindent
\global\let\lasttitle=N\global\everypar={}}%
\global\let\lasttitle=M%
\ignorespaces}
\def\abstract#1{\bgroup\petit\noindent
{\bf Abstract: }\ignorespaces#1\vskip28pt\egroup
\let\lasttitle=N
\parindent=0pt
\everypar={\global\parindent=\oldparindent
\global\let\lasttitle=N\global\everypar={}}%
\ignorespaces}
\def\titlea#1#2{\if N\lasttitle\else\vskip-28pt
     \fi
     \vskip18pt plus 4pt minus4pt
     \bgroup
\textfont0=\tbfontt \scriptfont0=\tbfonts \scriptscriptfont0=\tbfontss
\textfont1=\tbmt \scriptfont1=\tbms \scriptscriptfont1=\tbmss
\textfont2=\tbst \scriptfont2=\tbss \scriptscriptfont2=\tbsss
\textfont3=\tasys \scriptfont3=\tenex \scriptscriptfont3=\tenex
     \baselineskip=16pt
     \lineskip=0pt
     \pretolerance=10000
     \noindent
     \tbfontt
     \rightskip 0pt plus 6em
     \setbox0=\vbox{\vskip23pt\def\fonote##1{}%
     \noindent
     \if!#1!\ignorespaces#2
     \else\setbox0=\hbox{\ignorespaces#1\unskip\ }\hangindent=\wd0
     \hangafter=1\box0\ignorespaces#2\fi
     \vskip18pt}%
     \dimen0=\pagetotal\advance\dimen0 by-\pageshrink
     \ifdim\dimen0<\pagegoal
     \dimen0=\ht0\advance\dimen0 by\dp0\advance\dimen0 by
     3\normalbaselineskip
     \advance\dimen0 by\pagetotal
     \ifdim\dimen0>\pagegoal\eject\fi\fi
     \noindent
     \if!#1!\ignorespaces#2
     \else\setbox0=\hbox{\ignorespaces#1\unskip\ }\hangindent=\wd0
     \hangafter=1\box0\ignorespaces#2\fi
     \vskip18pt plus4pt minus4pt\egroup
     \nobreak
     \parindent=0pt
     \everypar={\global\parindent=\oldparindent
     \global\let\lasttitle=N\global\everypar={}}%
     \global\let\lasttitle=A%
     \ignorespaces}
 \def\titleb#1#2{\if N\lasttitle\else\vskip-28pt
     \fi
     \vskip18pt plus 4pt minus4pt
     \bgroup
\textfont0=\tenbf \scriptfont0=\sevenbf \scriptscriptfont0=\fivebf
\textfont1=\tams \scriptfont1=\tamss \scriptscriptfont1=\tbmss
     \lineskip=0pt
     \pretolerance=10000
     \noindent
     \bf
     \rightskip 0pt plus 6em
     \setbox0=\vbox{\vskip23pt\def\fonote##1{}%
     \noindent
     \if!#1!\ignorespaces#2
     \else\setbox0=\hbox{\ignorespaces#1\unskip\enspace}\hangindent=\wd0
     \hangafter=1\box0\ignorespaces#2\fi
     \vskip10pt}%
     \dimen0=\pagetotal\advance\dimen0 by-\pageshrink
     \ifdim\dimen0<\pagegoal
     \dimen0=\ht0\advance\dimen0 by\dp0\advance\dimen0 by
     3\normalbaselineskip
     \advance\dimen0 by\pagetotal
     \ifdim\dimen0>\pagegoal\eject\fi\fi
     \noindent
     \if!#1!\ignorespaces#2
     \else\setbox0=\hbox{\ignorespaces#1\unskip\enspace}\hangindent=\wd0
     \hangafter=1\box0\ignorespaces#2\fi
     \vskip8pt plus4pt minus4pt\egroup
     \nobreak
     \parindent=0pt
     \everypar={\global\parindent=\oldparindent
     \global\let\lasttitle=N\global\everypar={}}%
     \global\let\lasttitle=B%
     \ignorespaces}
 \def\titlec#1#2{\if N\lasttitle\else\vskip-23pt
     \fi
     \vskip18pt plus 4pt minus4pt
     \bgroup
\textfont0=\tenbfne \scriptfont0=\sevenbf \scriptscriptfont0=\fivebf
\textfont1=\tams \scriptfont1=\tamss \scriptscriptfont1=\tbmss
     \tenbfne
     \lineskip=0pt
     \pretolerance=10000
     \noindent
     \rightskip 0pt plus 6em
     \setbox0=\vbox{\vskip23pt\def\fonote##1{}%
     \noindent
     \if!#1!\ignorespaces#2
     \else\setbox0=\hbox{\ignorespaces#1\unskip\enspace}\hangindent=\wd0
     \hangafter=1\box0\ignorespaces#2\fi
     \vskip6pt}%
     \dimen0=\pagetotal\advance\dimen0 by-\pageshrink
     \ifdim\dimen0<\pagegoal
     \dimen0=\ht0\advance\dimen0 by\dp0\advance\dimen0 by
     2\normalbaselineskip
     \advance\dimen0 by\pagetotal
     \ifdim\dimen0>\pagegoal\eject\fi\fi
     \noindent
     \if!#1!\ignorespaces#2
     \else\setbox0=\hbox{\ignorespaces#1\unskip\enspace}\hangindent=\wd0
     \hangafter=1\box0\ignorespaces#2\fi
     \vskip6pt plus4pt minus4pt\egroup
     \nobreak
     \parindent=0pt
     \everypar={\global\parindent=\oldparindent
     \global\let\lasttitle=N\global\everypar={}}%
     \global\let\lasttitle=C%
     \ignorespaces}
 \def\titled#1{\if N\lasttitle\else\vskip-\baselineskip
     \fi
     \vskip12pt plus 4pt minus 4pt
     \bgroup
\textfont1=\tams \scriptfont1=\tamss \scriptscriptfont1=\tbmss
     \bf
     \noindent
     \ignorespaces#1\ \ignorespaces\egroup
     \ignorespaces}
\let\ts=\thinspace
\def\footnoterule{\kern-3pt\hrule width 2true cm\kern2.6pt}
\newcount\footcount \footcount=0
\def\advftncnt{\advance\footcount by1\global\footcount=\footcount}
\def\fonote#1{\advftncnt$^{\the\footcount}$\begingroup\petit
\parfillskip=0pt plus 1fil
\def\textindent##1{\hangindent0.5\oldparindent\noindent\hbox
to0.5\oldparindent{##1\hss}\ignorespaces}%
\vfootnote{$^{\the\footcount}$}{#1\vskip-9.69pt}\endgroup}
\def\item#1{\par\noindent
\hangindent6.5 mm\hangafter=0
\llap{#1\enspace}\ignorespaces}

\def\titleao#1{\vfill\supereject
\ifodd\pageno\else\null\vfill\supereject\fi
\let\header=N
     \bgroup
\textfont0=\tafontt \scriptfont0=\tafonts \scriptscriptfont0=\tafontss
\textfont1=\tamt \scriptfont1=\tams \scriptscriptfont1=\tamss
\textfont2=\tast \scriptfont2=\tass \scriptscriptfont2=\tasss
\textfont3=\tasyt \scriptfont3=\tasys \scriptscriptfont3=\tenex
     \baselineskip=18pt
     \lineskip=0pt
     \pretolerance=10000
     \tafontt
     \centerpar{#1}%
     \vskip75pt\egroup
     \nobreak
     \parindent=0pt
     \everypar={\global\parindent=\oldparindent
     \global\let\lasttitle=N\global\everypar={}}%
     \global\let\lasttitle=A%
     \ignorespaces}






\def\leaderfill{\kern0.5em\leaders\hbox to 0.5em{\hss.\hss}\hfill\kern
0.5em}
\newdimen\chapindent
\newdimen\sectindent
\newdimen\subsecindent
\newdimen\thousand
\setbox0=\hbox{\bf 10. }\chapindent=\wd0
\setbox0=\hbox{10.10 }\sectindent=\wd0
\setbox0=\hbox{10.10.1 }\subsecindent=\wd0
\setbox0=\hbox{\enspace 100}\thousand=\wd0
\def\contpart#1#2{\medskip\noindent
\vbox{\kern10pt\leftline{\textfont1=\tams
\scriptfont1=\tamss\scriptscriptfont1=\tbmss\bf
\advance\chapindent by\sectindent
\hbox to\chapindent{\ignorespaces#1\hss}\ignorespaces#2}\kern8pt}%
\let\lasttitle=Y\par}
\def\contcontribution#1#2{\if N\lasttitle\bigskip\fi
\let\lasttitle=N\line{{\textfont1=\tams
\scriptfont1=\tamss\scriptscriptfont1=\tbmss\bf#1}%
\if!#2!\hfill\else\leaderfill\hbox to\thousand{\hss#2}\fi}\par}
\def\conttitlea#1#2#3{\line{\hbox to
\chapindent{\strut\bf#1\hss}{\textfont1=\tams
\scriptfont1=\tamss\scriptscriptfont1=\tbmss\bf#2}%
\if!#3!\hfill\else\leaderfill\hbox to\thousand{\hss#3}\fi}\par}
\def\conttitleb#1#2#3{\line{\kern\chapindent\hbox
to\sectindent{\strut#1\hss}{#2}%
\if!#3!\hfill\else\leaderfill\hbox to\thousand{\hss#3}\fi}\par}
\def\conttitlec#1#2#3{\line{\kern\chapindent\kern\sectindent
\hbox to\subsecindent{\strut#1\hss}{#2}%
\if!#3!\hfill\else\leaderfill\hbox to\thousand{\hss#3}\fi}\par}
\long\def\lemma#1#2{\removelastskip\vskip\baselineskip\noindent{\tenbfne
Lemma\if!#1!\else\ #1\fi\ \ }{\it\ignorespaces#2}\vskip\baselineskip}
\long\def\proposition#1#2{\removelastskip\vskip\baselineskip\noindent{\tenbfne
Proposition\if!#1!\else\ #1\fi\ \ }{\it\ignorespaces#2}\vskip\baselineskip}
\long\def\theorem#1#2{\removelastskip\vskip\baselineskip\noindent{\tenbfne
Theorem\if!#1!\else\ #1\fi\ \ }{\it\ignorespaces#2}\vskip\baselineskip}
\long\def\corollary#1#2{\removelastskip\vskip\baselineskip\noindent{\tenbfne
Corollary\if!#1!\else\ #1\fi\ \ }{\it\ignorespaces#2}\vskip\baselineskip}
\long\def\example#1#2{\removelastskip\vskip\baselineskip\noindent{\tenbfne
Example\if!#1!\else\ #1\fi\ \ }\ignorespaces#2\vskip\baselineskip}
\long\def\exercise#1#2{\removelastskip\vskip\baselineskip\noindent{\tenbfne
Exercise\if!#1!\else\ #1\fi\ \ }\ignorespaces#2\vskip\baselineskip}
\long\def\problem#1#2{\removelastskip\vskip\baselineskip\noindent{\tenbfne
Problem\if!#1!\else\ #1\fi\ \ }\ignorespaces#2\vskip\baselineskip}
\long\def\solution#1#2{\removelastskip\vskip\baselineskip\noindent{\tenbfne
Solution\if!#1!\else\ #1\fi\ \ }\ignorespaces#2\vskip\baselineskip}


\long\def\definition#1#2{\removelastskip\vskip\baselineskip\noindent{\tenbfne
Definition\if!#1!\else\
#1\fi\ \ }\ignorespaces#2\vskip\baselineskip}
\def\frame#1{\bigskip\vbox{\hrule\hbox{\vrule\kern5pt
\vbox{\kern5pt\advance\hsize by-10.8pt
\centerline{\vbox{#1}}\kern5pt}\kern5pt\vrule}\hrule}\bigskip}
\def\frameddisplay#1#2{$$\vcenter{\hrule\hbox{\vrule\kern5pt
\vbox{\kern5pt\hbox{$\displaystyle#1$}%
\kern5pt}\kern5pt\vrule}\hrule}\eqno#2$$}
\def\typeset{\petit\noindent This book was processed by the author using
the \TeX\ macro package from Springer-Verlag.\par}
\outer\def\byebye{\bigskip\bigskip\typeset
\footcount=1\ifx\speciali\undefined\else
\loop\smallskip\noindent special character No\number\footcount:
\csname special\romannumeral\footcount\endcsname
\advance\footcount by 1\global\footcount=\footcount
\ifnum\footcount<11\repeat\fi
\gdef\leftheadline{\hbox to2.5true cc{\folio\hfil}\ignorespaces
\the\AUTHOR\unskip: \the\HEAD\hfill}\vfill\supereject\end}
